# Elastic higher-order topological insulator with topologically protected corner states

Haiyan Fan, Baizhan Xia*, Liang Tong, Shengjie Zheng, Dejie Yu

State Key Laboratory of Advanced Design and Manufacturing for Vehicle Body, Hunan University, Changsha, Hunan, People's Republic of China, 410082

**Topologically gapless edge states, characterized by topological invariants and Berry's phases of bulk energy bands, provide amazing techniques to robustly control the reflectionless propagation of electrons, photons and phonons. Recently, a new family of topological phases, dictated by the bulk polarization, has been observed, leading to the discovery of the higher-order topological insulators (HOTIs). So far, the HOTIs are only demonstrated in discrete mechanical and electromagnetic systems and electrical circuits with the quantized quadrupole polarization. Here, we realize the higher-order topological states in a two-dimensional (2D) continuous elastic system whose energy bands can be well described. We experimentally observe the gapped one-dimensional (1D) edge states, the trivially gapped zero-dimensional (0D) corner states and the topologically protected 0D corner states. Compared with the trivial corner modes, the topological ones, immunizing against defects, are robustly localized at the obtuse-angled but not the acute-angled corners. The topological shape-dependent corner states open a new route for the design of the topologically-protected but reconfigurable 0D local eigenmodes and provide an excellent platform for the topological transformation of elastic energy among 2D bulk, 1D edge and 0D corner modes.**

Topological states, theoretically predicted in condensed matter physics[1-3], have become the revolutionizing technologies for the robust control of the propagation of photons and phonons. The topological insulators (TIs) of photons and phonons were firstly realized in the broken time-reverse symmetry systems, in which the counterpropagating partners are perfectly prohibited[4-12]. Then, the quantum valley Hall effects[13-21] and the quantum spin Hall effects[22-28] were observed in time-reverse invariant photonic and phononic systems. These topological characteristics, dictated by integer topological invariants and Berry's phases of bulk energy bands, obey the bulk-edge correspondence principle, exhibiting the topologically protected gapless 1D edge states which immunize against large confined imperfectness and even spontaneous emissions. More recently, by gapping edge states, the higher-order topological phases, characterized by the bulk polarization, have been theoretically predicted[29-35]. And then, the HOTIs were experimentally implemented in mechanical[36], microwave[37], and optical[38] systems and electrical circuits[39].

The elastic wave propagation in well-structured phononic systems has exhibited exotic properties, from negative refraction[40], super focusing[41] to cloaking[42]. Inspired by the fascinating topological characteristics of acoustic and optical systems, the topologically protected edge states in elastic phononic plates[43-49] have been receiving great attentions and show excellent prospects for energy storing, information carrying and nondestructive testing. However, the HOTIs, going beyond the classical bulk-edge correspondence principle, have not been realized in elastic wave systems. Compared with optical and acoustic waves, the elastic waves consisting of in-plane and out-of-plane waves are the vector waves whose bulk polarizations are much more difficult to be well-controlled. In this paper, we design an elastic HOTI. The gapped 1D edge states, the trivially gapped 0D corner states and the topologically protected 0D corner states, instead of the gapless 1D edge states, are experimentally observed in the topologically nontrivial bandgap. The corner states, robustly localized at the corners of the elastic HOTI, not only seriously decay into the bulk, but also rapidly attenuate along edges. Differing from the previous HOTIs in square and cube lattices[29,35-37], the new developed topological corner states depend on both the bulk topology and the corner shapes. They can be pinned to the obtuse-angled corners valued by $2\pi/3$, but vanish at the acute-angled corners valued by $\pi/3$. This topological property provides an unprecedented opportunity for the topological design of elastic devices with reconfigurable local eigenmodes at corners.

A composite unit cell consisting of six nodes is presented in Fig. 1a. The continuous elastic honeycomb lattice formed by an array of composite unit cells (highlighted by a green hexagon) is depicted in Fig. 1b. This lattice is fabricated by cutting hexagonal blocks from an acrylic panel with the material properties of density $\rho=1190$kg/m$^3$, Poisson's ratio $v=0.35$, and Young's modulus $E=3.2$GPa. The thickness of the acrylic panel is $d=1.98$mm. The width and length of the acrylic beam are $w=5.02$mm and $L=15$mm, respectively. Two cylindrical nickel-plated neodymium magnets ($\rho=7400$kg/m$^3$, $v=0.28$, and $E=41$GPa), marked by light blue in Fig. 1a, are attached to the upper and lower sides of each node, working as additional masses. The height and radio of the magnet are $h=2.0$mm and $r=2.51$mm, respectively. The coupling within the composite unit cell is marked as the inter-cell coupling $l_{inter}$. The coupling among the neighboring unit cells is marked as the intra-cell coupling $l_{intra}$. The inter- and intra-cell couplings can be modulated by the lengths of acrylic beams.

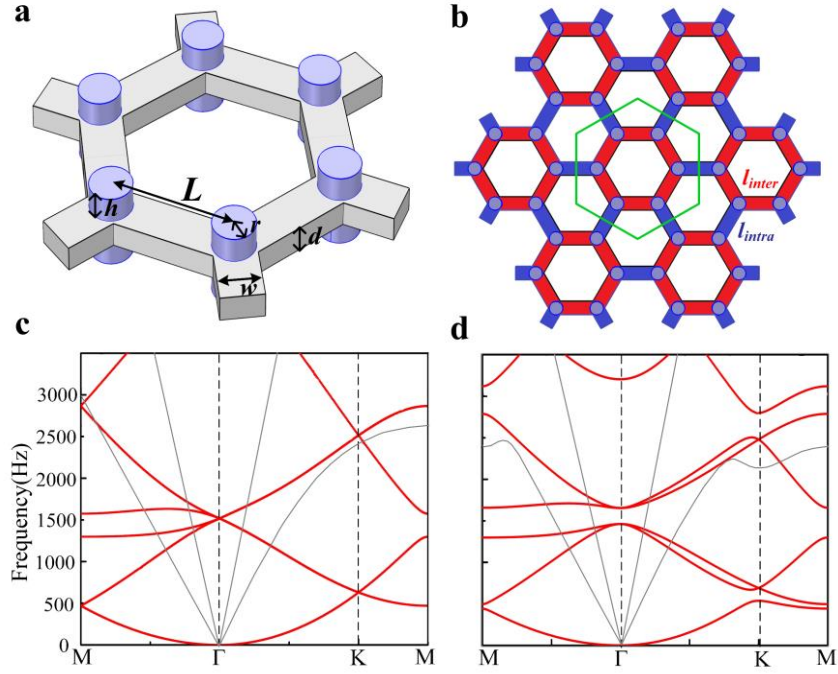

Fig. 1 | **a**, A composite unit cell of the honeycomb lattice, with six pairs of magnetic cylinders clamped at the six nodes. **b**, Partial selection of the lattice. Green hexagon delimits the composite unit cell. Red (blue) beams represent the inter-cell (intra-cell) beams whose coupling strength are defined by their length $l_{inter}$ ($l_{intra}$). **c**, Band structures of the composite unit cell with a double Dirac cone at a frequency of 1517Hz. Here, $l_{inter}$ equates to $l_{intra}$. **d,** Band structure with a complete bandgap in a shrunken lattice ($l_{inter}<l_{intra}$).

Under a long wavelength limit, this lattice can be approximated as a thin plate. The out-of-plane mode described by the parabolic dispersion is loosely coupled with in-plane modes, leading to that the out-of-plane bands (red lines) are identifiably uncoupled from the in-plane bands (gray lines). In this paper, the in-plane polarization which goes beyond the scope of this work is neglected. When $l_{inter}$ equates to $l_{intra}$, this elastic honeycomb lattice is a perfect honeycomb phononic crystal. As shown in Fig. 1c, there is a double Dirac cone, linearly degenerated by four bands, emerging at the Γ point of the irreducible Brillouin zone. This double Dirac cone is protected by the $C_{6v}$ symmetry of the lattice. If the mirror-reflection symmetry is broken, the quantum valley Hall effects, accompanied with the topologically protected gapless 1D edge states, can yield in this elastic honeycomb lattice[47]. Here, we will explore the higher-order topological phases which exhibit the gapped 1D edge states and the in-gap 0D corner states, instead of the gapless 1D edge states. Aiming as this issue, a new mechanism based on the modulation of the inter- and intra-cell couplings will be developed.

When $l_{inter}$ increases to $1.2L$, and at the same time, $l_{intra}$ reduces to $0.6L$, the double Dirac cone is lifted, yielding a complete bandgap ranging from 1503Hz to 1715Hz. On the contrary, when $l_{inter}$ reduces to $0.836L$ and $l_{intra}$ increase to $1.328L$, the double Dirac cone is also lifted for a bandgap ranging from 1461Hz to 1650Hz, inspected in Fig. 1d. Thus, when $l_{inter}<l_{intra}$ (referred to a shrunken lattice) and $l_{inter}>l_{intra}$ (referred to an expanded lattice), the elastic phononic plates support two new band diagrams, respectively. Although they are similar and cannot be intuitively distinguished, these two band diagrams essentially describe two distinct topological phases. Their displacement fields are inversed when crossing the gapless double Dirac point at $l_{inter}=l_{intra}$ (Seeing Supplementary Fig. S1), giving rise to a band inversion and a topological transition between trivial and nontrivial phases which can be demonstrated by the bulk polarization[29,30,35].

The symmetry-protected edge state on the boundary of a configuration is the fascinating property of the TI. In this study, we construct two different elastic phononic configurations with periodic boundaries along $x$ direction, but truncated along $y$. The first one is a ribbon configuration consisting of 9 expanded unit cells. Its band dispersion, presented in Fig. 2a, shows that there are only bulk bands which are separated by a complete bandgap, indicating that this expanded ribbon configuration is a trivial one without edge states. With the decrease of $l_{inter}/l_{intra}$, the bandgap will gradually decrease, and finally degenerate to a gapless one at $l_{inter}/l_{intra}=1$ (Seeing Supplementary Fig. S2). If $l_{inter}/l_{intra}$ is further reduced, the second ribbon configuration consisting of 9 shrunken unit cells is constructed. In this sample, the bandgap reopens, with two edges bands, denoted by the red lines in Fig. 2b. However, these 1D edge states are gapped, which is inherently different from the gapless 1D edge states protected by the time-reversal symmetry in the quantum Hall insulator.

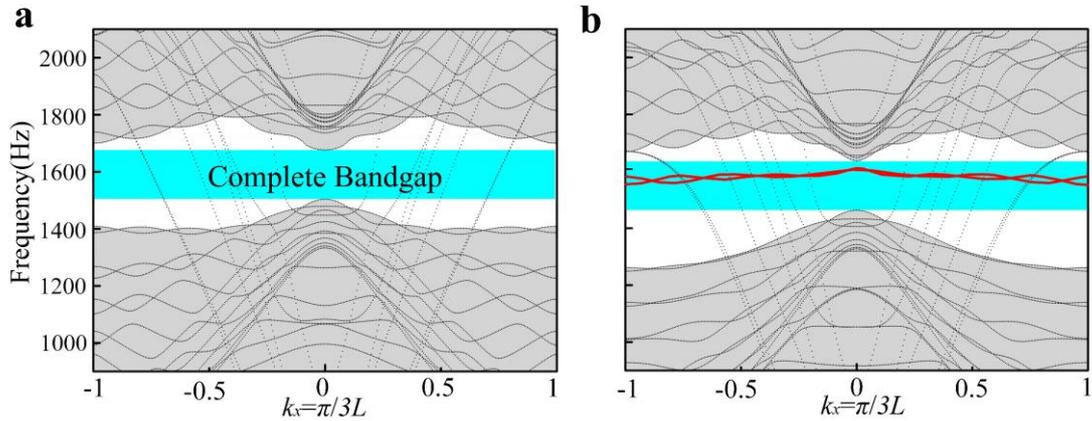

Fig. 2 | **a**, Band dispersion of a ribbon configuration consisting of an expanded phononic lattice with 9 cells along the *y* direction. **b**, Band dispersion of a ribbon configuration consisting of a shrunken phononic lattice with 9 cells along the *y* direction.

Three large experimental hexagon-shaped samples with 37 unit cells are depicted in Fig. 3a, 3b and 3c. The hexagon-shaped sample with expanded unit cells ($l_{inter}>l_{intra}$) distributing along the six boundaries is a trivial structure. Its numerically calculated eigenfrequencies presented in Fig. 3d show that there are only bulk modes, as expected. On the contrary, the second hexagon-shaped sample with shrunken unit cells ($l_{inter}<l_{intra}$) is a topological nontrivial structure. As presented in Fig. 3e, the gapped edge modes and the in-gap corner modes are generated in the topologically nontrivial bandgap between the lower- and the higher-frequency bulk modes. The simulated field profile presented in Fig. 3g show that for the gapped edge modes, the elastic wave energy is well localized along the boundaries of the topological hexagon-shaped sample, expect for the six corners. Besides of gapped edge modes, two classes of in-gap corner modes are in the bandgap. One is the topologically protected corner mode (marked by red in Fig. 3e) and the other one is the trivial corner mode (marked by blue in Fig. 3e). Their simulated field profile depicted in Fig. 3h (for a topological corner mode) and 3i (for a trivial corner mode) show that the elastic wave energy is strongly concentrated in the corners of the topological hexagon-shaped sample, which is essentially different from the edge mode (Fig. 3g) and the bulk mode (Fig. 3j). The nontrivial bulk polarization of a HOTI suggests that the topologically protected corner states exhibit a good immunity against defects. To confirm this, we deliberately design an imperfect hexagon-shaped sample by attaching one more magnet with the height of 1mm at the upper side of each node of 12 unit cells (marked by red dotted lines in Fig. 3c). Numerically evaluated eigenfrequencies presented in Fig. 3f show that the topological corner states (defined by red nodes) are well confined at the frequency around 1555Hz, perfectly isolated from the bulk and edge states, favorably evidencing the strong robustness of the topological corner states against moderate defects and disorders. On the contrary, the frequencies of trivial corner states (marked by blue nodes in Fig. 3f) are shifted from 1528Hz to 1476Hz, which reveals that the trivial corner states are very sensitive to defects and disorders.

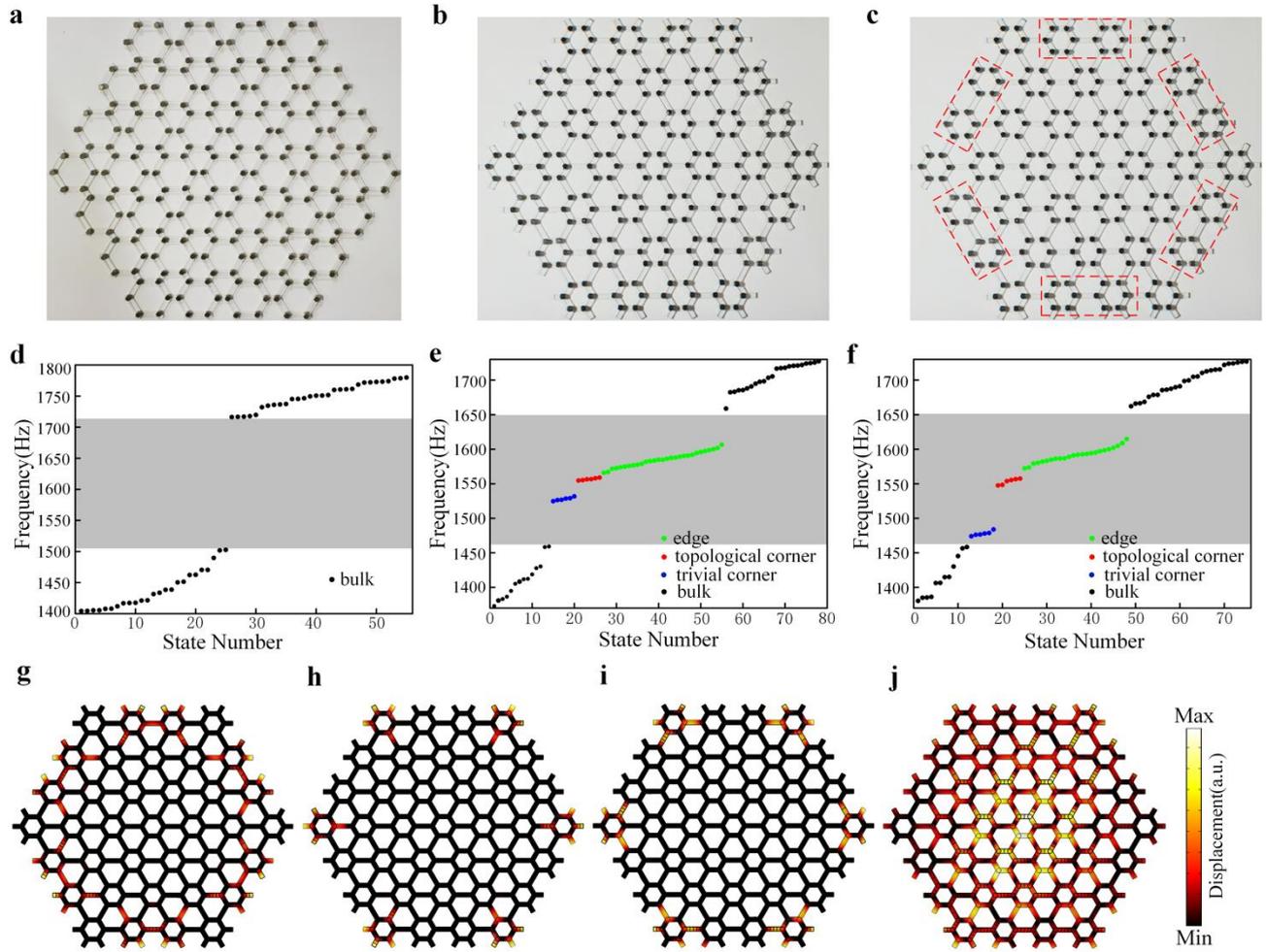

Fig. 3 | **a-c**, Images of hexagon-shaped samples with expanded unit cells, shrunken unit cells and defects (framed by red dashed lines) respectively. **d-f**. Numerically evaluated eigenfrequencies of hexagon-shaped samples with expanded unit cells, shrunken unit cells and defects. Green, red, blue and black dots denote gapped edge, topological corner, trivial corner and bulk modes, respectively. **g-j**, The simulated displacement field profiles of gapped edge, topological corner, trivial corner and bulk states, respectively.

The bulk, the edge and the corner transmission spectra of the trivial hexagon-shaped sample are presented in Fig. 4a. All measurements were performed by a scanning laser Doppler vibrometer (LV-S01), seeing Method for details. Within the complete bandgap, the bulk, the edge and the corner transmissions are very low, indicating that the propagation of elastic waves is efficiently blocked in this trivial sample. The bulk, the edge and the corner transmission spectra of the topological hexagon-shaped sample are presented in Fig. 4b. For the bulk transmission spectrum (the black curve), the peaks, separated by the complete bandgap, are observed in the lower- and the higher-frequency bulk regions. For the edge transmission spectrum (the green curve), a high peak, located in the bandgap, is observed around 1610Hz, being consistent with the edge mode presented in the simulations of Fig. 3e. The topological and the trivial corner states of this hexagon-shaped sample are very close to each other. As a result, the peaks, being consistent with both corner modes, are overlapped to be whole, expressing as a common peak around 1550Hz, as shown in the corner transmission spectrum (the red curve).

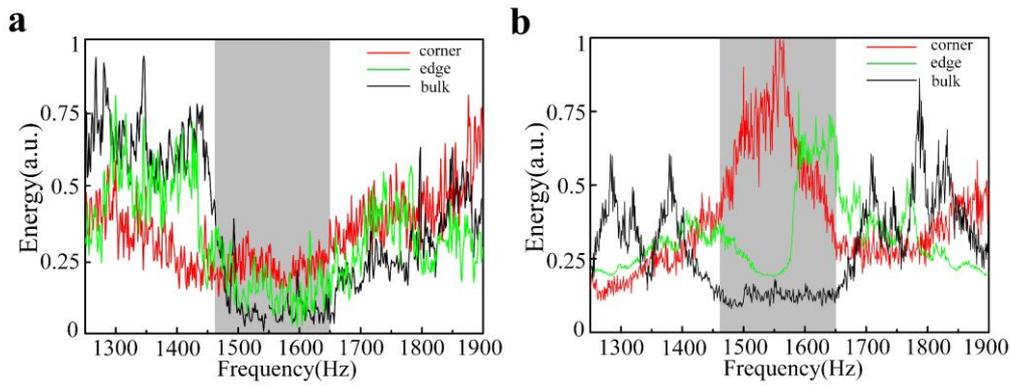

Fig. 4 | **a**, Measured bulk (black), edge (green) and corner (red) transmission spectra for a trivial hexagon-shaped sample. **b**, Measured bulk (black), edge (green) and corner (red) transmission spectra for a topological hexagon-shaped sample.

A large triangular-shaped sample with shrunken unit cells ($l_{\text{inter}} > l_{\text{intra}}$) distributing along three boundaries is depicted in Fig. 5a. Its numerically evaluated eigenfrequencies presented in Fig. 5c shows that this triangular-shaped sample has bulk, gapped edge and gapped corner modes. The simulated field profiles presented in Fig. 5e-5g show that for the above eigenmodes, the elastic wave energy is localized along the boundaries (except for the three corners), at the corners and within the bulk of the triangular-shaped sample, respectively. When the perfection of the triangular-shaped sample is broken by attaching one more magnet with a height of 1mm on the upper side of each node of 6 unit cells (marked by red dotted line in Fig. 5b), the gapped corner modes will shift to the lower frequency region and even move out of the bandgap (Fig. 5d), indicating that these trivially gapped corner states are very sensitive to defects. The topologically protected corner mode immunizing against defects is not observed in this triangular-shaped sample, which is inherently distinct from the hexagon-shaped sample. The experimentally measured corner transmission spectra depicted in Fig. 5h show that the peak for the perfect triangular-shaped sample is around 1500Hz, being consistent with its trivially gapped corner modes. However, the peak for the imperfect sample is shifted to 1450Hz, due to the variation of the trivially gapped corner modes.

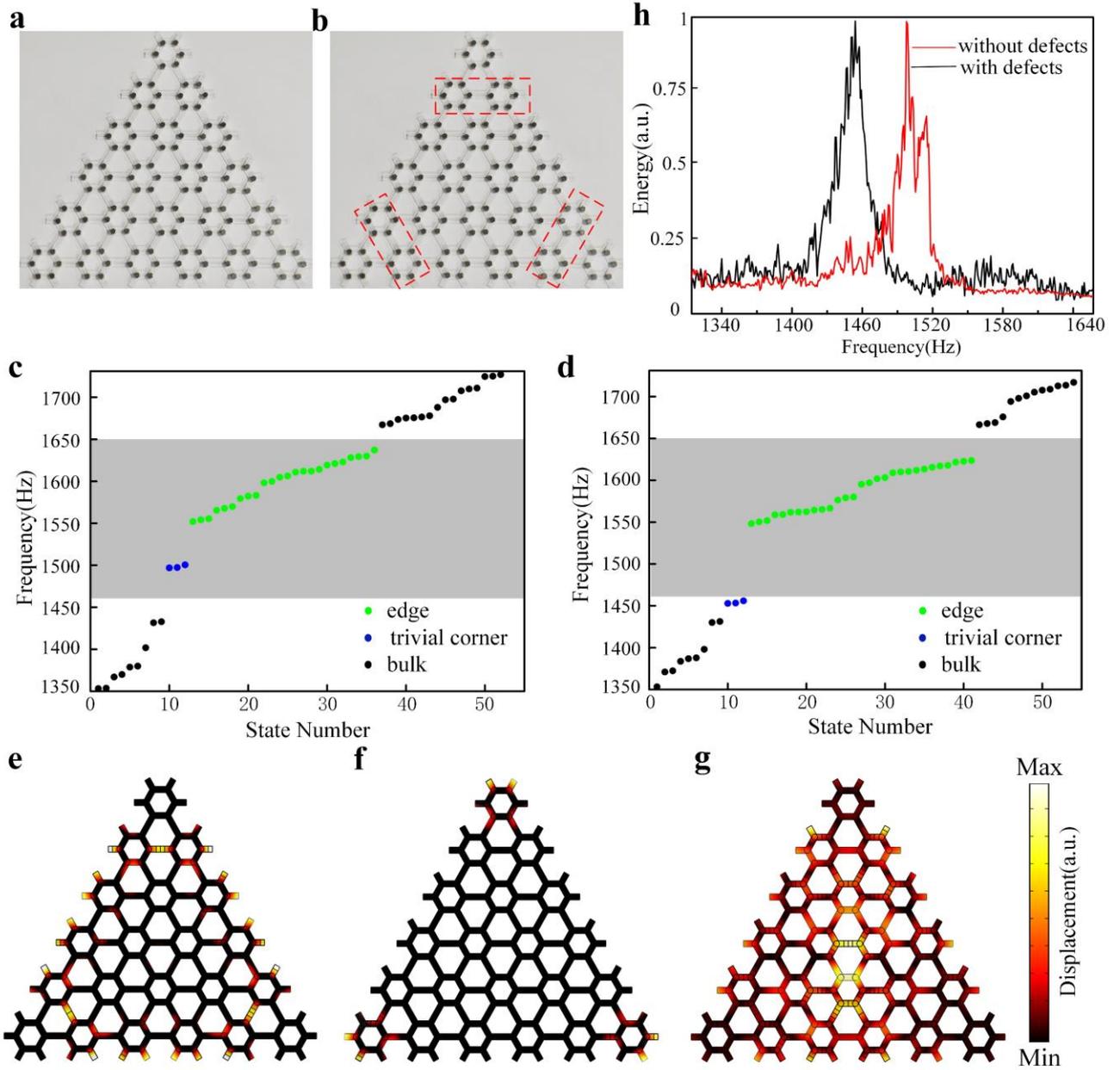

Fig. 5 | **a-b**, Images of the triangular-shaped samples without and with defects, respectively. The defect nodes are framed by red dashed lines. **c-d**, Numerically evaluated eigenfrequencies for triangular-shaped samples without and with defects. Green, blue and black dots denote gapped edge, gapped corner and bulk modes, respectively. **e-g**, The simulated displacement field profiles of edge, corner and bulk states, respectively. **h**, Measured corner transmission spectra for the hexagon-shaped samples with defects (black curve) and without defects (red curve).

According to the different corner states of the hexagon- and the triangular-shaped samples, a physical consequence of our elastic HOTI is that the topologically protected corner modes only emerge at the obtuse-angled corners valued by $2\pi/3$, but not the acute-angled corners valued by $\pi/3$. To explain this phenomenon, a topological index $N=|N_+-N_-|$, capturing the interplay between the topology of the bulk Hamiltonian and the topological structure of the defect, is introduced. $N=|N_+-N_-|$ is the number of stable modes bounded to corners[50,51]. $N_+$ and $N_-$ are respectively the numbers of zero modes with topological charges +1 and −1. When the intra-cell coupling is larger than the inter-cell coupling which can be neglected in its limiting case, the topological index $N$ can be determined by the number of zero modes at corners. At the acute-angled corner valued by $\pi/3$, there are four zero modes. Two of them have the topological charge +1 and the other two possess the topological charge −1. Thus, the topological index $N$ is $N=N_+-N_-=2-2=0$. As a result, there are no stable modes localized at the acute-angled corners, as shown in Fig. 5c. At the obtuse-angled corner valued by $2\pi/3$, there are three zero modes. Two of them have the same charge and the other one locating at the sharpest corner possesses the opposite charge. Thus, the topological index $N$ is $N=|N_+-N_-|=|2-1|=1$, assuring that the stable modes will be localized at the obtuse-angled corners, as presented in Fig. 3e. The topological parallelogram-shaped sample simultaneously possessing the obtuse- and the acute-angled corners (seeing the Supplementary Figure S3) further verifies the topological shape-dependent corner states.

In conclusion, this work develops an elastic HOTI by modulating the inter- and intra-cell couplings in a honeycomb lattice. The topologically protected 0D corner modes, lying in the nontrivial bandgap, can robustly concentrate the wave energy at the obtuse-angled corners valued by $2\pi/3$, but not the acute-angled corners valued by $\pi/3$. Based on this shape-dependent characteristic, our elastic HOTI provides an additional degree-of-freedom, which is apart from the bulk polarization, to turn on/off the topologically protected local eigenmodes at corners. It is believed that our work exhibits a great ability to control the elastic wave propagation in an unprecedented way and provides an excellent platform to design the new type of elastic topological devices which can topologically transform the elastic wave energy among the bulk, the edge and the corner modes. Furthermore, our physical mechanism to realize the elastic higher-order topological phase is simple, so it can be directly extended to the three-dimensional (3D) HOTIs and even Weyl semimetals, exhibiting in-gap hinge states over a wide range of disciplines, including optics, acoustics and mechanical vibrations.

## Acknowledgements

This work is supported by the Foundation for Innovative Research Groups of the National Natural Science Foundation of China (Grant No. 51621004) and the Joint Fundation for Equipment Pre-research by Ministry of Education of China (6141A02033216).

## Author contributions

H.Y. and B.Z. designed the experiments. H.Y. and T. performed the experiments. H.Y. and S.J. carried out the numerical simulations. H.Y., B.Z. and D.J. wrote the manuscript. B.Z supervised the project. All the authors contributed to the analyses of the results.

## Competing interests

The authors declare no competing interests.

## Methods

**Simulation.**

All numerical simulations were performed by COMSOL Multiphysics, Solid Mechanics modules, commercial software based on finite element method (www.comsol.com). In the numerical simulations of Figs. 3d-f, we used a large system with 169 composite unit cells for a much clearer separation of boundary modes. In the numerical simulations of Figs. 5c and 5d, we used a large system with 91 composite unit cells for a much clearer separation of boundary modes.

**Signal measurements.**

Our samples were made of the transparent square acrylic plate with the thickness of 1.98mm, by using a laser cutting technique. For the vibration excitation, a vibration exciter (HEV-20) was used. The diameter of the ejector rod of the exciter was 4 mm. The sharp ejector rod was pressed on the surface of the panel. The displacement field was scanned by a laser Doppler vibrometer (LV-S01). As the laser beam of the vibrometer was perpendicular to the panel, only the vertical displacement component (namely the out-of-plane wave energy) was captured. The displacement signal was recorded by a LMS SCADAS Mobile System.

For the hexagon-shaped samples, we measured the response of the lower right corner by exciting and measuring the same nodes. The transmission spectra shown as the red curves in Fig. 4 were the average values of the results of the three nodes of the lower right corner. We then measured the edge transmission (the green curve in Fig. 4) by exciting and measuring the nodes located at the bottom edge. Finally, we measured the bulk transmission spectra (the black curve in Fig. 4) by exciting and measuring the nodes located in the bulk. For the triangular-shaped sample, we measured the response of the lower right corner by exciting and measuring the same nodes. The transmission spectra (the red and the black curves in Fig. 5h) were the average values of the results of the two nodes of the lower right corner. The $y$-axis of Figs. 4 and 5h was the mechanical energy $\varepsilon \propto \Delta z^2$, where $\Delta z$ was the measured amplitude of the out-of-plane vibration.